# XML Data Integrity Based on Concatenated Hash Function

Baolong Liu
School of Computing & Engineering
University of Huddersfield
Huddersfield, UK

Joan Lu
School of Computing & Engineering
University of Huddersfield
Huddersfield, UK

Jim Yip
School of Computing & Engineering
University of Huddersfield
Huddersfield, UK

*Abstract*—**Data integrity is the fundamental for data authentication. A major problem for XML data authentication is that signed XML data can be copied to another document but still keep signature valid. This is caused by XML data integrity protecting. Through investigation, the paper discovered that besides data content integrity, XML data integrity should also protect element location information, and context referential integrity under fine-grained security situation. The aim of this paper is to propose a model for XML data integrity considering XML data features. The paper presents an XML data integrity model named as CSR (content integrity, structure integrity, context referential integrity) based on a concatenated hash function. XML data content integrity is ensured using an iterative hash process, structure integrity is protected by hashing an absolute path string from root node, and context referential integrity is ensured by protecting context-related elements. Presented XML data integrity model can satisfy integrity requirements under situation of fine-grained security, and compatible with XML signature. Through evaluation, the integrity model presented has a higher efficiency on digest value-generation than the Merkle hash tree-based integrity model for XML data.**

## I. INTRODUCTION

General applications of data integrity could exist in many domains, including e-government, e-commerce, e-financial services, e-business, e-banking, e-learning, e-healthcare, mobile communications, heterogeneous networks, digital factories, multi-agent systems, and grid computing [1-14]. For example, Wu and Chen described the need for data integrity when official documents are being transmitted between government agencies for e-government in Taiwan [1, 2]. O'Neill pointed out the importance of data integrity through an assessment of a bank's web service [6]. IBM gives an example of data integrity as follows: Assume the data is a funds transfer and the hacker alters a random piece of the data that happens to be the account number. When the bank decrypts the data, the account number is not a valid account; therefore, the data tampering is detected and the transaction is not completed. However, assume instead that the data altered by the hacker node is the amount of money and, changed it from 1000 units to 9000 units [15]. In this case, the transaction would be completed using the incorrect amount. Therefore, research into this area would be of great benefit.

There are two approaches to ensuring integrity for XML data. The first tries to add additional elements to XML data to record the integrity information [16, 17]. Without cryptography, this method is easily attacked by a hacker. The second approach is based on a cryptography mechanism, and adopts a hash function to ensure integrity [18, 19, 20, 21, 22]. This cryptography-based approach has a higher security level than the first approach, although there are two major shortcomings in this kind of integrity model for XML data.

Firstly, existing integrity models generate a digest value for XML data content without considering XML data features. For non-XML data formats, a user can directly generate digest value of the data content to ensure integrity, but protecting data content integrity alone is not enough for XML data. For example, a signed XML data can be copied to another document but still keep signature valid. This problem can be utilized by an attacker to forge a document with a valid signature. Therefore, besides data content integrity, XML data integrity should also consider element location information and element context meaning under a fine-grained security situation. In this paper, fine-grained security means that users would encrypt or sign portions of XML data rather than the whole. Location information of an XML element refers to the position of this element in the XML data [17]. An element has an entire meaning related to its position in XML data, and will lose original meaning if the position has been changed. Thus, XML data integrity should also protect location information of an XML element in XML data. Another factor which affects the meaning of XML elements is the context relationship. For example, the element will no longer have its original meaning without context relationship in an XML data, and the paper defines this as context referential integrity, in other words, an XML element has an entire meaning only related to other elements in the same XML data, but there is no mechanism which can be used to protect this meaning in an existing integrity model for XML data.

Secondly, most of these models are based on the Merkle hash tree [20, 21], when generate digest value, the Merkle hash tree will increase virtual nodes. The hash times will also be increased because of these virtual nodes, and this leads to a low efficiency on digest value-generation.

Motivated by the problems above, this paper aims to present XML data integrity requirements combined with XML data features. Based on the XML data integrity requirements presented, it proposes an integrity model for XML data, and improves the efficiency of digest value-generation for XML data.

This paper presents an XML data integrity model named as CSR. The model consists of three parts, and CSR is an acronym for these parts: 'C' for content integrity, 'S' for structure integrity, and 'R' for context referential integrity.

The three parts are combined with a concatenated hash function. Content integrity is used to ensure XML data content integrity by using a concatenated hash function. Structure integrity is used to protect the location information of an element in XML data by hashing an absolute path string from the root node. Finally, context referential integrity protects the integrity of context-related elements. This paper also describes the combination of the model with XML specification, and integrates the integrity model presented into the XML signature. From evaluation, the integrity model presented has a higher efficiency on digest value-generation than the Merkle hash tree-based integrity model for XML data.

### A. Contribution

The major contribution of this paper is the XML data integrity requirement considering XML data features, and satisfies the requirements with an integrity model for XML data with a higher efficiency. The detail is as follows.

- This paper gives a description of XML data integrity requirements related to XML data features under fine-grained XML security. Three aspects considered are content integrity, structure integrity, and context referential integrity.

- Based on presented requirements, an integrity model for XML data has been built based on concatenated hash function. As far as is known by the author, this is the first attempt to give a formal integrity model for XML data considering XML data features.

- Based on a concatenated hash function to generate digest value for XML data, this method has a higher efficiency than the Merkle hash tree-based digest value-generation process.

### B. Structure of the paper

The remainder of this paper is organized into sections. Section 2 describes the related work of XML data integrity, and section 3 introduces theory guidance of this research. Section 4 gives XML data integrity requirement and the model definition for XML data integrity. Section 5 presents the experimental results. Section 6 makes a comparison of existing XML integrity approaches, and section 7 concludes the paper.

## II. RELATED WORK

Through investigation, there are two approaches to ensure integrity for XML data. The first tries to enter additional elements in XML data to record the integrity information. Hussain maintained the integrity of XML signatures using the manifest element [16]. McIntosh presented an element position attack, and solved this problem by adding additional objects in XML data [17]. Without cryptography, this kind of method is easily attacked by a hacker.

The second approach is based on a cryptography mechanism, and adopts a hash function to ensure integrity. DOM-HASH is the first algorithm proposed by Maruyama to calculate a hash value for XML data [18]. In this algorithm, MD5 and SHA1 were adopted to generate hash values with four different node types related to XML data. The four node typed include element, attribute, processing instruction (PIs), and text. This algorithm is limited to the contents of the XML data and, therefore, does not provide for authentication of the internal or external subset of the DTD. Inspired by DOM-HASH, the XHASH algorithm has been proposed by Brown [19]. The XHASH makes use of two parameters: the first is the digest function such as SHA1; the second, which is optional, can be used to determine how non-significant space characters will be handled by default. However, possible values for this attribute are limited to 'default' and 'preserved'. Thus, there is no known way to explicitly specify that non-significant space characters should be discarded. W3C published XML signature specifications in 2000 (Second Edition in 2008) [23, 24]. This specification provides the format for data integrity expressions in XML signatures, and gives the optional algorithm to generate digest value, such as SHA-1, SHA-256. However, signed XML data can be copied to another document but still keep signature valid. Devanbu adopted DOM-HASH and the Merkle hash function to maintain the integrity of XML data queries [20]. Bertino also adopted the Merkle hash tree to handle XML documents [21]. These two methods provide a solution to generate digest values of XML data based on the Merkle hash tree. However, the element's attribute integrity has been ignored in the model presented by Bertino [25]. Furthermore, the Merkle hash tree has a low efficiency. Qiao presented a united-message digest method related to XML data integrity [22]. Based on cryptography, this kind of approach has a higher security level than the first approach. But, this kind of approach still has some problems described above.

## III. THEORY GUIDANCE

### A. Types of integrity mechanisms

In order to ensure integrity, there are perfect means to assure the information integrity, such as hashes or check-sum mechanisms [26]. Both are used to detect changes to the original information. However hashes are more focused on malicious changes while check-sums are applied to detect coincidental changes.

In this paper, data integrity is ensured by a hash function mechanism. The reasons for adopting a hash function as an integrity method is as follows [26].

- A checksum is useful in detecting accidental modification such as corruption to stored data or errors in a communication channel.

- Checksums provide no security against a malicious agent as their simple mathematical structure makes them easy to break. An example is CRC series.

- A hash function has one-way and collision-resistant features with a complex mathematical model, it provides a higher level security compared to a check-sum.

### B. Theory basis

The integrity model in this paper referred to the model presented by DOM-HASH and Bertino although the construction process is different. The integrity model presented by Bertino is based on Merkle hash tree [21]. In this paper, the integrity model CSR is constructed based on the theory of the concatenated hash function. Just like the Merkle hash tree, the concatenated hash function also is designed to handle tree structure hash process. The reasons for adopting a concatenated hash function to construct the integrity model for XML data is as follows.

- Concatenated hash functions can handle arbitrary tree structure, but the Merkle hash function mainly deals with binary tree structure [27]. Thus, a concatenated hash function is more suitable to handle XML data.

- Concatenated hash functions can decrease the numbers of hash processes, thus it has higher efficiency in digest value-generation for XML data than the Merkle hash tree.

## IV. XML DATA INTEGRITY MODEL CSR

### A. XML data integrity requirement

In order to illustrate the requirement of XML data integrity, an example has been given in Fig. 1, and it is a real application document derived from website of NPL [28]. Note that some details have been omitted.

```
01 <Certificate>
02   <Title>Certificate of calibration</Title>
03   <ReferenceNumber>TDFRG</ReferenceNumber>
04   <Description>A single-mode Fibre Attention
                  Standard...</Description>
05   <Data>This reported expanded uncertainty is based
              on...</Data>
06   <Measurements>
07     <Description>The measurement of the spectral
                    attenuation...</Description>
08     <Table>Designed figure used in measurement</Table>
09   </Measurements>
10   <Results>
11     <Description>The total attenuation...</Description>
12     <Graph>Chart related to measurement results</Graph >
13     <Table>Figure of measurement results</Table>
14     <Results>
       ⋮
15 </Certificate>
```

Figure 1. A certificate for fault detection

- Content integrity (CI)

The XML data content refers to element name, attribute, and values of an element or sub XML data. Content integrity means that XML data content will not be changed or destroyed in transmitting or storage. This is ensured by generating a digest value of XML data. As shown in Fig. 1, content integrity for element 'Title' should include tag name 'Title' and related value 'Certificate of calibration'.

- Data structure integrity (STI)

An XML data structure integrity protects the location information of an element in XML data [17]. It means that if the location of an element in the XML data has been changed, it will lead to an invalid verification. Location information of an XML element refers to the position of this element in the XML data. Element location information consists of three parts: parent, level, and order in sibling. This position helps people to understand the meaning of the element. In other words, an element will have different meanings when it is located in different positions in XML data. As shown in Fig. 1, there are three 'Description' elements in line 04, 07, 11. The 'Description' element has a completely different meaning related to its location: line 04 is a description for certificate information; line 07 is a description for measurement; line 11 is the description for measured results. Thus, location information for an XML element is an important aspect and needs to be protected.

- Context referential integrity (CRI)

In most situations, when adopting XML data format, without considering element context relationship, only one element will also lose its original meaning. For example, as shown in Fig. 1, the measurement result has a completely meaning related to measurement method or technique deployed in the certificate. These two elements are generated by different responsibilities. So, it can not be signed by only one person, or signed together, because each unit is only responsible for its own role. Under this situation, element 'Certificate/Results' has a completely meaning that is only related to element 'Certificate/Measurements'. It means this kind of testing results occurrence corresponds to a given measurement. In other words, an XML element has an entire meaning only when related to other elements in the same XML data, and these elements have been defined as context-related elements in this paper.

Context referential integrity is used to protect context-related elements of an element in XML data. It will provide a binding between an element and context-related elements. This means if context-related elements of an element have been altered, it will also lead to an invalid verification.

In summary, the basic requirement for XML data integrity is that XML data has not been changed, destroyed, or lost in an unauthorized or accidental manner. Considering the features of XML data as analyzed above, the detailed integrity requirements for XML data are as follows.

- XML data content, including element name, value, and attribute, has not been changed, destroyed, or lost.

- Element location information, including element parent, level, and order in sibling, should be protected in an XML data.

- In order to ensure a completely meaning of an element in an XML data, context-related elements should be protected together with this element.

### B. Definition of integrity model CSR

To develop a formal model for XML data integrity this paper introduces a definition for XML data as in definition 4.1. Based on the requirement for XML data integrity presented above and XML data definition, the integrity model CSR for XML data is defined as follows:

**Definition 4.1** An XML data is a tuple $X_D = (V, r, E_d, \phi E_d)$ [21], where:

- $V = V^e \cup V^a$ is a set of nodes representing elements and attributes, respectively. Each $v \in V^a$ has an associated value $val \in Value$; each $v \in V^e$ may have an associated data content.

- $r$ is a node representing the document element ( Called XML data root);

- $E_d \subseteq V_d \times V_d$ is the set of edges.

- $\phi E_d$ is the edge labelling function.

**Definition 4.2** Content integrity $CI(v)$

XML content integrity should protect name, attributes, value of an element or sub XML data. Let $X_D$ be an element or sub XML data, and $h$ be a collision-resistant one way hash function. "$\|$" denotes the concatenation operator. The $CI(v)$ associated with $X_D$ is a function, and for each, $v \in V$

$$CI(v) = \begin{cases} h(((v.content) \,\|\, (v.attribute) \,\|\, (CI(v.child1) \,\|\, \cdots \,\|\, CI(v.childn))) & \text{if } v \text{ is a vertice} \\ h((v.content) \,\|\, (v.attribute)) & \text{if } v \text{ is a leafnode} \end{cases}$$

(1)

Formula (1) only provides the digest value for an element or portions of XML data. Here, $v.content = v.name \cup v.value$. This is also ensured structures of element or sub XML data, because $v.content$ should include element name and related value. Through the properties $v.content$ and $v.attribute$, XML data content integrity is ensured. The definition is also based on a concatenated hash function, meaning that all children of an element are concatenated together before, generating a digest value.

**Definition 4.3** Label for an XML node $L(v)$

$L(v) = C_1 C_2$, here, $C_1 \in$ Integer is represented by the level of corresponding node $v$. $C_2 = sibling(v)$ represents the order of sibling nodes, and $sibling(v)$ is the function to get sibling order of node $v$.

For example, the label for element "Certificate\Results\Description" in the Fig. 1 is expressed as: $L(\text{Certificate} \setminus \text{Results} \setminus \text{Description}) = 31$

**Definition 4.4** Structure integrity $ST(v)$

For each $v \in V$, $ST(v) = h(path(r, v))$  (2)

Here, the result is the digest value of path string related to $v \in V$, and $r$ is the root of XML data. $p = path(r, v)$: $p \in string$, denotes a path from root $r$ to current element $v$. The result $p$ is an ordered sequence of one or more nodes $p \in r_{L(r)} / v^1_{L(v^1)} / \cdots / v^m_{L(v^m)} / v_{L(v)}$, and $r$ is the root of XML data, $v^1$ is the child of node $r$, $v^m$ is the child of $v^{m-1}$, and $v$ is the current element. $L(v)$ is the label for an internal node.

Through definition 4.3 and 4.4, the location of an element can be expressed as a path string from root node to current node. This path records the level, sibling order, and parent of an element. Through the digest value of this path string, element location information would be protected.

**Definition 4.5** Context referential integrity $CRI(v)$

Suppose $w$ is the context-related element of an XML data $v$, and write as $v \to w$, then, $CRI(v) = h(CI(w) \| ST(w))$  (3)

Formula (3) is the context referential integrity in XML data, here $w \in V$. This definition includes integrity of context-related element content and its location information. Context-related elements can be chosen by a signer before signing an XML data with considering context relationship.

Based on above definition, the model CSR is defined in definition 4.6.

**Definition 4.6** Definition of Integrity model CSR
$CSR(v) = h(CI(v) \| ST(v) \| CRI(v))$  (4)

The result of formula (4) is a digest value for the XML data. This value consists of three parts: $CI(v)$, $ST(v)$, and $CRI(v)$, with the three parts combined by a concatenated hash function. Here, $v \in V$ is the node set of the XML data. $CI(v)$ is a digest value of an element or sub XML data, used to protect the XML data content. $ST(v)$ is a digest value of element location information, used to protect the position of an element or sub XML data in the XML data. $CRI(v)$ is a digest value of context-related elements, used to protect context relationship of an element. $h$ is a collision-resistant one-way hash function. The combination of these three parts is by string concatenation, i.e., by hashing the concatenated string $x_1 \| \ldots \| x_l$.

In case an element copied from an XML data to another document which has the same structure as original one, the original data creation timestamp is used to distinguish them as defined in definition 4.7. This definition is a combination of timestamp with integrity model CSR.

**Definition 4.7** Let $S = h(t \| CSR(v))$ be the digest value that is finally signed. Here, $t$ is an attribute of the creation timestamp related to XML data $X_D$.

### C. Integrity analysis

The integrity proofs are expressed by three theorems. Theorem 4.1 provides the evidence of structure integrity, theorem 4.2 proves context referential integrity, and theorem 4.3 proves that a signed XML data can not be copied into another document.

**Theorem 4.1** If an element $v \in V$ in XML data $X_D$ and $X_D^{'}$, and $X_D \neq X_D^{'}$, without considering context-related elements, then $CSR(v) \neq CSR^{'}(v)$.

**Proof:** In this theorem, because $v$ is the same in XML data $X_D$ and $X_D^{'}$, and without considering context-related elements, there is the same $CI(v)$, $CRI(v)$ in $X_D$ and $X_D^{'}$. If $CSR(v) \neq CSR^{'}(v)$, there must be different $ST(v)$ in $X_D$ and $X_D^{'}$. In other words, $v$ has different location in $X_D$ and $X_D^{'}$. Location information consists of three parts: parent, level, and order of sibling.

Assume the path from root to current element $v$ in XML data $X_D$ is: $p_1 = v_{11} / v_{2j} / \ldots / v_{ij}, i, j \in \text{int}$

Assume the path from root to current element $v$ in XML data $X_D^{'}$ is: $p_2 = r_{11} / r_{2n} / \ldots / r_{mn}, m, n \in \text{int}$

The value of $ST(v)$ in XML data $X_D$:

$ST(v) = h(path(p_1)) = h(v_{11} / v_{2j} / \ldots / v_{ij})$

The value of $ST(v)$ in XML data $X_D^{'}$:

$ST^{'}(v) = h(path(p_2)) = h(r_{11} / r_{2n} / \ldots / r_{mn})$

Because $X_D \neq X_D^{'}$, there are two kinds of situations as follows:

- Different level

If $v$ has different level in XML data $X_D$ and $X_D^{'}$, then $i \neq m$, and $h(v_{11}/v_{2j}/.../v_{ij}) \neq h(r_{11}/r_{2n}/.../r_{mn})$. Thus, $CSR(v) \neq CSR^{'}(v)$. It also means element $v$ has different ancestors.

- Different sibling order

If $v$ has different sibling order in XML data $X_D$ and $X_D^{'}$, then $j \neq n$, and $h(v_{11}/v_{2j}/.../v_{ij}) \neq h(r_{11}/r_{2n}/.../r_{mn})$.

Thus, $CSR(v) \neq CSR^{'}(v)$.

The theorem 4.1 is used to judge the data integrity by using XML data structure. Because the two XML data have different structures, the element location will be changed, thus, from the defined integrity model, they will have different digest values, and lead to an invalid verification.

**Theorem 4.2** An element $v \in V$ in XML data $X_D$ and $X_D^{'}$, if the context-related element is $T_1$ in XML data $X_D$, $T_1^{'}$ in XML data $X_D^{'}$, and $T_1 \neq T_1^{'}$, then $CSR(v) \neq CSR^{'}(v)$.

**Proof:** If $X_D \neq X_D^{'}$, from theorem 4.1, then $CSR(v) \neq CSR^{'}(v)$

If $X_D = X_D^{'}$ and $T_1 \neq T_1^{'}$, then the value of $CSR(v)$ in XML data $X_D$ is expressed as follows:

$CSR(v) = ST(v) \| SE(v) \| CRI(v) = ST(v) \| (CI(T_1) \| ST(T_1)) \| CRI(v)$

The value of $CSR^{'}(v)$ in XML data $X_D^{'}$:

$CSR^{'}(v) = ST(v) \| SE(v) \| CRI(v) = ST(v) \| (CI(T_1^{'}) \| ST(T_1^{'})) \| CRI(v)$

$T_1 \neq T_1^{'}$ means $T_1, T_1^{'}$ have different content, or different structure.

If $T_1, T_1^{'}$ have different content, then $CI(T_1) \neq CI(T_1^{'})$ Thus, $CSR(v) \neq CSR^{'}(v)$

If $T_1, T_1^{'}$ have different structure, then $CI(T_1) \neq CI(T_1^{'})$ and $ST(T_1) \neq ST(T_1^{'})$ Thus, $CSR(v) \neq CSR^{'}(v)$.

The theorem 4.2 is used to check for changes in context-related elements. If the same element has different context-related elements, regardless of whether or not the two XML data have the same structure, it will lead to an invalid verification.

**Theorem 4.3** An element $v \in V$ in XML data $X_D$, if $X_D$ is signed and copied to another XML data $X_D^{'}$, it will lead to an invalid verification.

**Proof:** If $X_D$ and $X_D^{'}$ have not same structure and content, then from theorem 4.1, there has $CSR(v) \neq CSR^{'}(v)$. It will lead to an invalid verification.

In many cases, if two XML data have same structure and content, they should be the same XML data. Thus, an element copied from one XML data to another will not affect the validation result. However, XML data has its own creating time, which can be used to judge the validation of an element in an XML data. Thus, the integrity model combined with timestamp, to prevent an elements being copied maliciously from one XML data to another.

Assume $S(v)$ is the signature related to element $v$, therefore: the value of $S(v)$ in XML data $X_D$:

$S(v) = h(h(t_1) \| CSR(v)))$

The value of $S(v)$ in $X_D^{'}$: $S(v) = h(h(t_2) \| CSR(v)))$

If $X_D$ and $X_D^{'}$ have a different creation time, therefore $h(t_1) \neq h(t_2)$, and it will lead to an invalid verification. If $X_D, X_D^{'}$ have a same creation time, and $X_D$ has the same structure and content as $X_D^{'}$, this means that $X_D$ is the same XML data as $X_D^{'}$.

### D. Efficiency analysis

The cost of a CSR model consists of the following two factors: the node size and the depth size. In a $k-ary$ tree with a depth of $m$, in the worst situation, then number of nodes that could be hashed is $N = \sum_{x=1}^{m} k^{x-1} = \frac{k^m - 1}{k-1}$, and the number of hash required $W = \sum_{x=1}^{m} x k^{x-1} = \frac{mk^{k+1} - (m+1)k^m + 1}{(k-1)^2}$.

The time complexity of an iterative hash function $h$ can be described as a function of its input size $l$ by the function, $T(l) = c_1 \left( \left\lfloor \frac{l}{D} \right\rfloor + 1 \right) + c_2$, where $D$ is constant [29]. If $v$ is a vertex of XML data $X_D$, $in\deg(v)$ denote the depth of vertex $v$, that is the number of predecessors of $v$ in $X_D$. Let $S$ be a subtree of $X_D$. The two components of the integrity cost for $S$ are defined as follows. The node size $S_n$ of $S$ is the number of its vertices. The depth size $S_d$ of $S$ is the sum of the depth of its vertices, that is $S_d = \sum_{v \in S} in\deg(v)$. Then, the rehashing overhead is given by a linear combination of the node size and the depth size of $S$, that is $c \mid v \mid + c^{'} \sum_{v \in S} in\deg(v) = cS_n + c^{'}S_d$, where both $c$ and $c^{'}$ are constants. The verification time is a quantity of the form $c \mid v \mid + c^{'} \sum_{v \in S} in\deg(v)$.

### E. Combination with XML specification

XML security research has two sides: firstly, how traditional security technologies can be used to solve problems existing in XML data; secondly, how to describe the security technologies in XML format. Based on the theory model presented for XML data integrity, this sub section describes how the theory model is expressed in XML format. The XML data content integrity has been described in the XML signature specification by W3C, thus, this section only gives the description for structure integrity, and context referential integrity.

#### 1) Specification for structure integrity

The structure integrity is ensured by three elements as follows.

- The 'STIGenerate Algorithm' is an element, which describes the algorithm used to generate the location information of an element in the original XML data.

- The content of the 'DigestMethod' element is the definition of digest algorithm adopted in this specification, and the default algorithm is SHA-1.

- The content of the 'DigestValue' element shall be the base64 encoding of this bit string viewed as a 20-octet octet stream.

```
<STI name="structure integrity" xmlns="http://www.example.org">
  <STIGenerate Algorithm="http://www.example.org/xmldsig-
      csr/#STI" />
  <DigestMethod
      Algorithm="http://www.w3.org/2000/09/xmldsig#sha1"/>
  <DigestValue>49-2A-ED-1A-5A-E1-BD-9C-59-04-19-58-8F-B7-
      08-5C-19-14-15-11</DigestValue>
</STI>
```

Figure 2. Example of structure integrity

An example of structure integrity is shown in Fig. 2.

Syntax: Schema for STI is shown in Fig. 3.

```
<?xml version = "1.0" encoding = "UTF-8"?>
<xsd:schema xmlns:xsd = "http://www.w3.org/2001/XMLSchema"
    elementFormDefault = "qualified">
  <xsd:element name = "STI" type = "STIType"/>
  <xsd:complexType name = "STIType" mixed = "true">
    <xsd:sequence>
      <xsd:element ref = "STIGenerate"/>
      <xsd:element ref = "DigestMethod"/>
      <xsd:element ref = "DigestValue"/>
    </xsd:sequence>
  </xsd:complexType>
  <xsd:element name = "STIGenerate">
    <xsd:complexType>
      <xsd:attribute name = "Algorithm" use = "optional" type =
          "xsd:anyURI"/>
    </xsd:complexType>
  </xsd:element>
  <xsd:element name = "DigestMethod">
    <xsd:complexType>
      <xsd:attribute name = "Algorithm" use = "optional" type =
          "xsd:anyURI"/>
    </xsd:complexType>
  </xsd:element>
  <xsd:element name = "DigestValue" type = "xsd:string"/>
</xsd:schema>
```

Figure 3. Schema for STI

### 2) Specification for context referential integrity

In specification, context referential integrity includes four elements as follows:

- The 'CRIGenerate Algorithm' is an element, which describes the algorithm used to generate the digest value of context-related elements.

- The content of the 'RelatedNode' is an element, which is used to record the context-related elements.

- The content of the 'DigestMethod' element is the definition of digest algorithm adopted in this specification, and the default algorithm is SHA-1.

- The content of the 'DigestValue' element shall be the base64 encoding of this bit string viewed as a 20-octet octet stream.

An example of CRI is shown in Fig. 4.

```
<CRI name="referential integrity"
        xmlns="http://www.example.org">
  <CRIGenerate Algorithm="http://www.example.org/xmldsig-
      csr/#CRI"/>
  <RelatedNode>#myDate</RelatedNode>
  <DigestMethod
      Algorithm="http://www.w3.org/2000/09/xmldsig#sha1"/>
  <DigestValue>36-C3-C5-A4-02-41-A9-0F-38-B7-C1-7C-7A-
      A0-A5-DE-7D-3A-75-9</DigestValue>
</CRI>
```

Figure 4. Example of CRI

Syntax: Schema for CRI is shown in Fig. 5.

```
<?xml version = "1.0" encoding = "UTF-8"?>
<xsd:schema xmlns:xsd = "http://www.w3.org/2001/XMLSchema"
    elementFormDefault = "qualified">
  <xsd:element name = "CRI" type = "CRIType"/>
  <xsd:complexType name = "CRIType" mixed = "true">
    <xsd:sequence>
      <xsd:element ref = "CRIGenerate"/>
      <xsd:element ref = "RelatedNode"/>
      <xsd:element ref = "DigestMethod"/>
      <xsd:element ref = "DigestValue"/>
    </xsd:sequence>
  </xsd:complexType>
  <xsd:element name = "CRIGenerate">
    <xsd:complexType>
      <xsd:attribute name = "Algorithm" use = "optional"  type =
          "xsd:anyURI"/>
    </xsd:complexType>
  </xsd:element>
  <xsd:element name = "RelatedNode" type = "xsd:string"/>
    <xsd:element name = "DigestMethod">
    <xsd:complexType>
      <xsd:attribute name = "Algorithm" use = "optional" type =
          "xsd:anyURI"/>
    </xsd:complexType>
  </xsd:element>
  <xsd:element name = "DigestValue" type = "xsd:string"/>
</xsd:schema>
```

Figure 5. Schema for CRI

## V. PERFORMANCE EVALUATION

### A. Evaluation environment

All testing was performed on a PC with a 2.39 GHz Pentium (R) 4 processor, 0.99GB of RAM, and MS Windows XP operating system. The programming language is Microsoft C#.

### B. Evaluation results

The integrity model for XML data presented by Bertino is based on the Merkle hash tree. The model CSR in this paper is based on a concatenated hash function. DOM-HASH is also based on an iterative algorithm, thus, when all of them have the same node size, the efficiency depends on the depth of XML data. There are five element numbers on every level in this testing. An evaluation has been made to compare the efficiency between these models. Let $H_i, i \in \text{int}$ be the depth of XML data, and the time requirement expressed as $T(H_i), i \in \text{int}$ . The comparison has been made based for two different hash algorithms, sha-1 and sha256 as shown in Fig. 6 and Fig. 7.

data width, the model CSR also is the most efficiency than others model as shown in Fig. 8 and Fig. 9. Compared to Fig. 6 and Fig. 7, XML data depth has a significant effect on XML data integrity generation process.

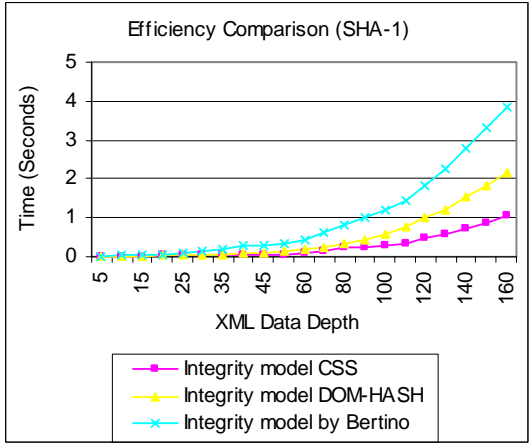

Figure 6. Efficiency comparison based on SHA-1 for XML data depth

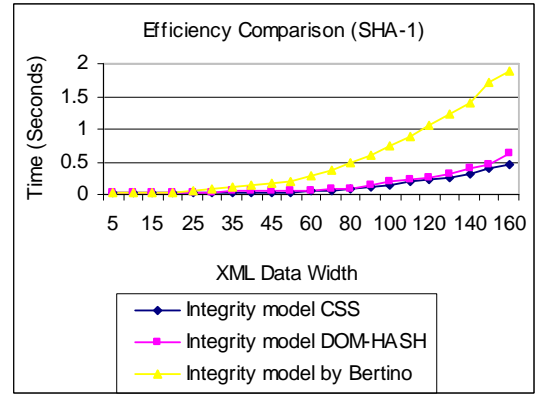

Figure 8. Efficiency comparison based on SHA-1 for XML data width

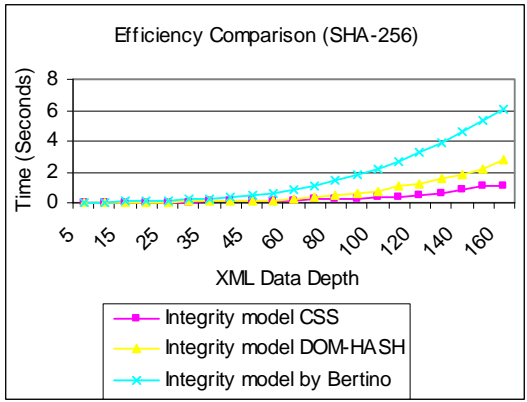

Figure 7. Efficiency comparison based on SHA-256 for XML data depth

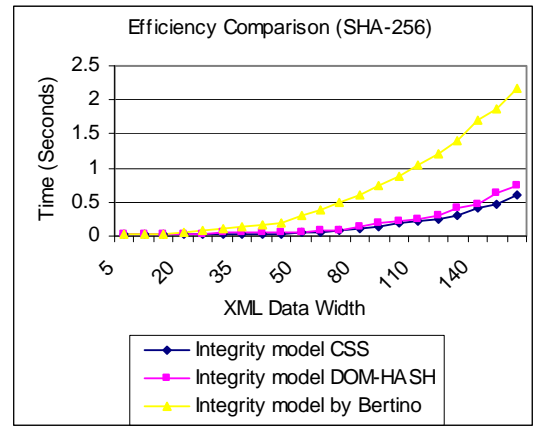

Figure 9. Efficiency comparison based on SHA-256 for XML data width

Fig. 6 shows that, these models have almost the same efficiency when XML data depth is less than 30. When the XML data depth is increased, the concatenated hash function-based integrity model CSR has the highest efficiency compared to integrity model DOM-HASH and integrity model by Bertino. The integrity model DOM-HASH has a higher efficiency when compared to integrity model by Bertino, and this is obvious when XML data has a higher depth. It can be calculated that the model CSR has 49.03% higher efficiency than DOM-HASH, and 74.72% higher efficiency than the integrity model by Bertino. Fig. 7 has the same development trend as Fig. 6, but because the algorithm sha256 is slower than sha-1, the total time overhead is increased as shown in Fig. 7. This indicates that although different hash algorithms have an impact on efficiency, the integrity model CSR presented is still the most efficient under different hash algorithms, and this is determined by integrity model mechanism, having nothing to do with adopted hash algorithms.

Without changing node size and numbers, when these nodes are at the same level, this report defined it as XML

The cause of this result is the different numbers of hash computations in the three models. Fig. 10 shows the total hash times of the three integrity models used in the testing sample. Bertino's model hashes the leaf node with $h(h(v.val) \| h(v.name))$, and there are 8 hash processes for each element, thus, with the increasing XML data depth, the element also increased as shown in Fig. 10. DOM-HASH hashes the leaf node with $h(v.elem \| v.text \| v.pi)$, and there is only 1 hash process for each element. In model CSR, the leaf node returned directly with 1 hash process, and the non-leaf node will have 2 hash process. Based on an iterative hash function, the integrity model hash of the leaf node will increase the hash times, which will lead to a low efficiency. Based on the concatenated hash function, the integrity model CSR concatenates the child node first, and then generates a digest value. Because of decreased the hash times, the integrity model CSR presented has a higher efficiency than other integrity models.

TABLE 1.    COMPARISON OF EXISTING INTEGRITY MODELS

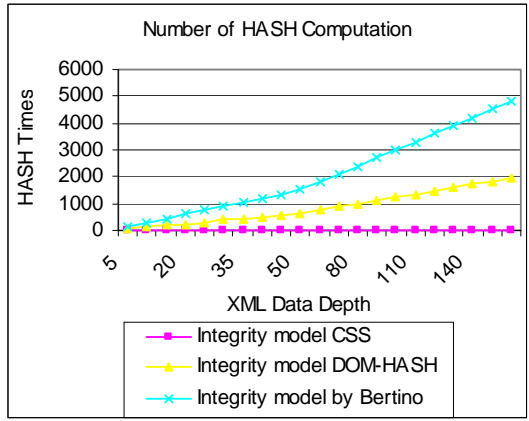

Figure 10. Comparison for numbers of hash computations

## VI.  DISCUSSION & ANALYSIS

### A.  Existing integrity approach comparison

In order to summarize the advantages of the XML data integrity model CSR presented, this paper makes a comparison of existing integrity models as shown in table 1.

Through comparison, the similarities of the integrity model CSR compared to existing models mainly focus on two aspects.

- The integrity model CSR also adopts a bottom-up iterative hash process as with DOM-HASH, Devanbu's, and Bertino's integrity model.

- The integrity model DOM-HASH, XHASH, Devanbu's model, Qiao's model, and model CSR ensure element name, attribute, and value, except Bertino's model ignored the attribute integrity of an element.

As shown in table 1, only the integrity model CSR for XML data provides overall integrity aspects, including data content, element location information, and element context meaning. Based on this comparison, the major differences of the model presented compared to others is as follows.

- Integrity model CSR is a formal model for XML data considering XML data features.

DOM-HASH and XHASH just consider the hash objectives, and the model by Devanbu and Bertino focus on the digest value-generation process. The model CSR combined the XML data features, such as the element location and context-related elements for example.

| Model Name | Description | Theory basis | Hash times each process |
|---|---|---|---|
| DOM-HASH by Maruyama [18] | $dos(v) = h(v.elem \| v.text \| v.pi)$<br>$\{Re\ s(v) = h(h(v.attr) \| dos(v) \| dos(v.child1) \| dos(v.sibling1) \| ... \| dos(v.siblingn))$<br>Here, $v$ is the element set of XML data, $h$ is a collision-resistant one-way hash function. | Iterative hash function | 1 or 2 |
| XHASH by Brown [19] | $dos(v) = h(v.elem \| v.text \| v.pi)$<br>$\{Re\ s(v,s) = h(h(v.attr) \| dos(v) \| dos(v.child1) \| dos(v.sibling1) \| ... \| dos(v.siblingn))$<br>Here, $v$ is the element set of XML data, $h$ is a collision-resistant one-way hash function. $s$ is default processing of non-significant SPACE characters. | Iterative hash function | 1 or 2 |
| XML Data integrity by Devanbu [20] | $f(v) = \{\begin{array}{l} h(v) \\ h(v, f(v_1), f(v_2), \cdots, f(v_k) \end{array}$<br>Here $v$ is a sink node, $v_1 \cdots v_k$ are the successor of $v$. $h$ is a collision-resistant one-way hash function. | Merkle hash function and DOM-HASH | 1 |
| XML Data integrity by Bertino [21] | $MhXd(v) = \{\begin{array}{l} h(h(v.val) \| h(v.name)) \quad if \quad v \in V_d^s \\ h(h(v.content) \| h(v.tagname) \| MhXd(child(1,v)) \| ... \| MhXd(child(n,v))) \quad if \quad v \in V_d^s \end{array}$<br>Here, $v$ is the element set of XML data, $h$ is a collision-resistant one-way hash function. | Merkle hash function | 3 |
| XML Data integrity by Hussain [16] | <Manifest> contains the data whose location is going to change and apply an XSLT transform to omit the URI attributes | N/A | N/A |
| XML Data integrity by Qiao [22] | $\{\begin{array}{l} Info(BCD \ldots M \ldots) = h(Info(A-B)), \cdots, H(Info(A-M)), \cdots \\ U-digest: h(Info(BCD \ldots M \ldots)) \end{array}$<br>Here, Info(A-B), …,Info(A-M), … is the sub XML data, Info(BCD…M…) is the united hashed result, and $h$ is a collision-resistant one-way hash function. | N/A | N/A |
| XML Data integrity model CSR | $CSR(v) = h(CI(v) \| ST(v) \| CRI(v))$ , here $v$ is the element set of XML data, $h$ is a collision-resistant one-way hash function. $CI(v)$ is the content integrity of signed elements, $ST(v)$ is the structure integrity, and $CRI(v)$ is the context referential integrity. | Concatenated hash function | 1 |

- Integrity model CSR not only ensure the integrity of data content, but also provides a method for digest value-generation process

The integrity model DOM-HASH and XHASH just provide the integrity objects which include element name,

attribute, and value, without describing the process of digest value-generation process. The integrity model CSR not only ensures the integrity of data content, but also describes the digest value-generation process. Two kinds of element have been involved, the leaf node and vertices. It will directly return the concatenation of content and attribute if the node is the leaf node, otherwise it will iteratively call the function.

- Bertino's model ignored the integrity attribute.

The content integrity in Bertino's model is only from $h(h(v.val) \parallel h(v.name))$. This formula does not consider the integrity attribute. In integrity model CSR, the integrity content includes $v.content \parallel v.attribute$, and $v.content = v.name \cup v.value$.

- Different hash times in the mathematical model.

Based on the Merkle hash function, hashing the leaf node will increase virtual nodes, and then increase the node numbers which need to be hashed, which can lead to a low efficiency. Based on concatenated hash function, this paper concatenates the child node firstly, and then generates a digest value. It has been proved that increasing hash numbers will not improve the security of hash function [30]. Thus, the model presented has the same security level as DOM-HASH and Bertino's integrity model, but because of decreased hash times, the presented hash process has a higher efficiency.

*B. Compatibility with XML signature specification*

The "XML signature Syntax and Processing" recommendation is an internet standard which defines syntax and processing model of a special format for digital signatures [23]. Standardized contents describe clear statement of the regulations on XML signature to maximize the security and the extent of the standardized contents, integrity, message and user authentication and non-repudiation. These signatures are represented in an XML format and can sign arbitrary resources, including XML and parts thereof.

The structure and processing of XML signatures introduces some interesting concepts which will be explained briefly. The primary elements of XML signatures are digital signature information and digest value information (The presentation of XML schema is as shown in Fig. 11). Signature elements consist of "SignedInfo" with digital signature information, "SignatureValue" with actual digital signature value and "KeyInfo" with digital signature key information. In particular, "SignedInfo" describes how signature information is standardized, the algorithm for the signature and the subordinate algorithm. "Reference" consists "DigestMethod", the algorithm summarizing signature data, and the element "DigestValue" showing the result. "KeyInfo" described in XML security is used to illustrate key information in XML digital signature.

```
<Signature>
  <SignedInfo>
    <CanonicalizationMethod/>
    <SignatureMethod/>
    <Reference>
      <DigestMethod/>
      <DigestValue/>
    </Reference>
  </SignedInfo>
  <SignatureValue>
  <KeyInfo>
</Signature>
```

Figure 11. XML digital signature elements

As described in our proposed scheme, the result of CSR is a hash value, thus, it can be described in element "DigestValue". The model CSR illustrates the method which is used to generate digest value, and it can be described in element "DigestMethod". Context-related elements can be described in "Transforms" element. Therefore, the proposed XML data integrity model CSR is compatible with the XML signature specification, and can be used in XML signature.

## VII. CONCLUSION AND FUTURE WORK

This paper discovered new integrity features related to XML data. Based on the XML data integrity requirement presented, the paper proposed a formal integrity model CSR for XML data. The paper also improved the digest value-generation process for XML data using a concatenated hash function. This paper draws the following conclusions.

- XML data integrity should not only consider data content integrity, but also needs to protect element location information and element context meaning under a fine-grained security situation.

- The improved digest value-generation process for XML data has a higher efficiency than the Merkle hash function-based fingerprint generation process.

- The formal integrity model CSR for XML data satisfies the XML data integrity requirements, and is compatible with XML signature.

Generally, XML data pass a hierarchical network of responsibilities (e.g. employees, supervisors) with different roles, thus it needs complex workflow on an XML data with multiple signatures based on element's context meaning presented in this paper. In order to enhance the constraints for workflow, a new multi-signer scheme has to be created. Future work identified below mainly focuses on XML multi-signature based on presented XML data integrity model, to build a constraint among multi-signers. The details is as follows.

- Build an algorithm to divide the XML data according to different responsibilities.

- Propose the constraint among multi-signers, this constraint being based on a hierarchical network.

- Integrate presented XML data integrity approach into the scheme of XML multi-signature.